\documentclass[a4paper,11pt]{article}
\pdfoutput=1 

\usepackage{jinstpub} 
\newcommand{\angstrom}{\textup{\AA}}

\title{An innovative technique for TPB deposition on convex window photomultiplier tubes}

\author[a]{M.~Bonesini,}
\author[b,c]{T.~Cervi,}
\author[c]{A.~Falcone,}
\author[d]{U.~Kose,}
\author[a]{R.~Mazza,}
\author[b,c]{A.~Menegolli,}
\author[c]{C.~Montanari,}
\author[d]{M.~Nessi,}
\author[c]{M.C.~Prata,}
\author[c]{A.~Rappoldi,}
\author[c]{G.L.~Raselli,} 
\author[c]{M.~Rossella,}
\author[b,c,1]{M.~Spanu\note{Corresponding author. Now at Brookhaven National Laboratory, Brookhaven NY, USA.},}
\author[a,e]{M.~Torti,}
\author[d]{W.~Vollenberg}
\author[d]{and A.~Zani}

\affiliation[a]{INFN - Sezione di Milano Bicocca, Piazza della Scienza 3, Milan, Italy}
\affiliation[b]{University of Pavia, Via Agostino Bassi 6, Pavia, Italy}
\affiliation[c]{INFN - Sezione di Pavia, Via Agostino Bassi 6, Pavia, Italy}
\affiliation[d]{CERN, Espl. des Particules 1, Geneva 23, Switzerland}
\affiliation[e]{University of Milano Bicocca, Piazza della Scienza 3, Milan, Italy}

\emailAdd{maura.spanu@cern.ch}

\abstract {
Tetraphenyl-butadiene (TPB) is an organic fluorescent chemical compound generally used as wavelength shifter thanks to its extremely high efficiency to convert ultra-violet photons into visible light. 
A common method to use TPB with detectors sensitive to visible light, such as photomultiplier tubes
(PMTs),
is to deposit thin layers on the device window.\\
To obtain effective TPB layers, different procedures can be used.
In this work a specific evaporation technique adopted to coat 8~in convex windows photomultiplier tubes is presented. It consists in evaporating TPB by means of a \textit{Knudsen cell}, which allows to strictly control the process,
and in a rotating sample support, which guarantees the uniformity of the deposition.\\
Simulation results and experimental tests demonstrate the effectiveness of this evaporation technique from the point of view of
deposition uniformity and light conversion efficiency.}  


\begin{document}
\maketitle


\section{Introduction}

Measurement of scintillation light in liquefied noble gases plays
an important role in many experiments dedicated to neutrino
physics and Dark Matter research. Photomultiplier tubes (PMTs)
presently represent the preferred readout devices to collect the
emitted light in large volume detectors, which are needed for rare
events physics. Various experiments already employ, or are foreseen to use, large area
PMTs directly immersed in liquid argon (LAr) at cryogenic
temperature. This is the case of ICARUS T600~\cite{ICARUS}, SBND~\cite{SBND}
and protoDUNE-Dual Phase~\cite{protoDUNE} with, respectively,
360, 120 and 36 8 in 
HAMAMATSU R5912-MOD PMTs. \\
The glass window of this PMT model is not transparent to the scintillation light produced by liquid argon ($\lambda=128$~nm).
Anyway, the PMTs sensitivity to vacuum-ultra-violet (VUV) photons can be achieved by depositing a layer of a proper wavelength shifter on the PMT window. \\
Tetraphenyl-Butadiene (TPB) is the most common wavelength shifter used in LAr detectors, due to its extremely high efficiency to convert VUV light into visible photons~\cite{Jerry}.
This paper presents the results of a careful study on a specific technique allowing to obtain a thin 
and uniform coating of TPB on the convex window of 8~in PMTs.
The technique consists in evaporating TPB by means of a  \textit{Knudsen cell}, which allows to strictly control the process, and a rotating PMT holding support,
which guarantees the uniformity of density of the deposition.
A detailed analysis of the deposited densities as a function of the position and
for different evaporation conditions was carried out using numerical simulations of the evaporation process and by means of experimental tests. Finally, the evaporation quality was validated in terms of light yield conversion, by using an optical instrumentation designed for quantum efficiency (QE) measurements
in the VUV range.

\section{TPB evaporation process}

\begin{figure}
	\center
	\includegraphics[scale=0.3]{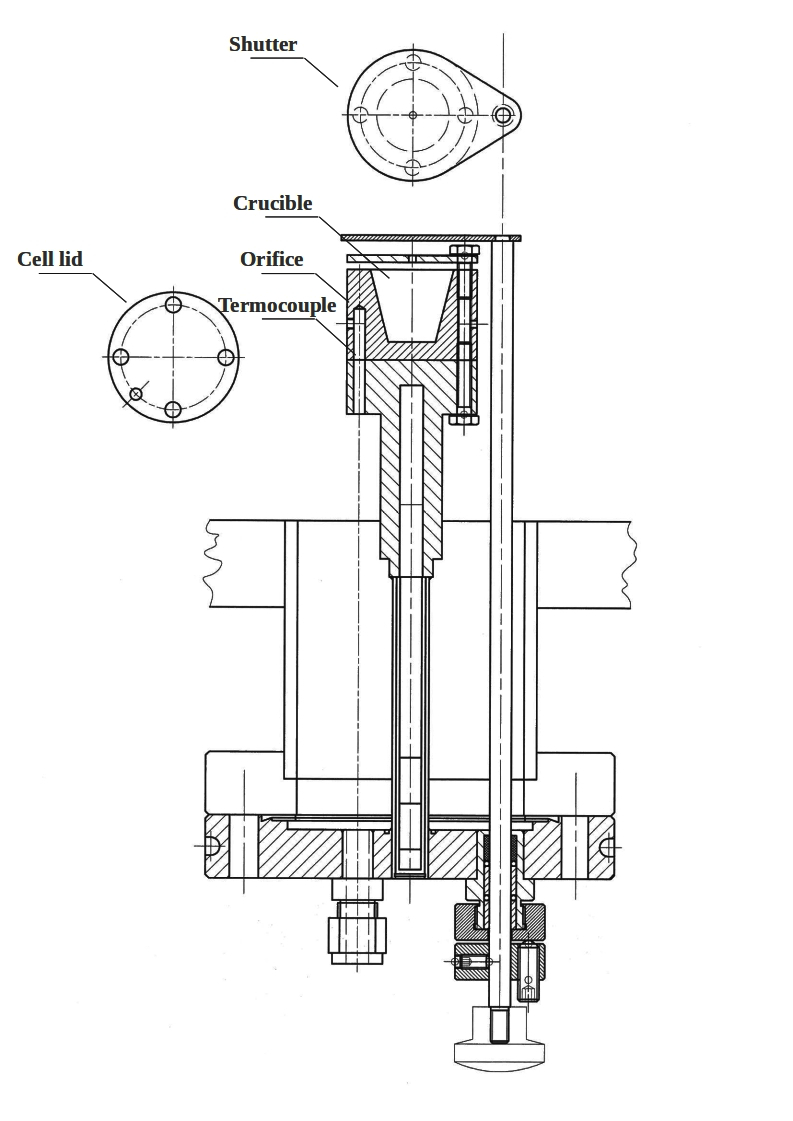}
	\caption{Schematic drawing of the adopted Knudsen cell.} \label{K-cell}
\end{figure}

1,1,4,4-Tetraphenyl-1,3-butadiene (TPB) is an organic chemical compound with formula 

\[
(C_{6}H_{5})_{2} C=CHCH=C(C_{6}H_{5})_{2}.
\] 
By visual inspection it appears composed by white and yellow-white needles. TPB absorbs light on a broad wavelength range, from the deep UV (VUV) to its emission peak, which lies at approximately $425 \pm 50$~nm. Scintillation-grade TPB, which is used in the detection of LAr scintillation light, has a purity greater than $> 99\%$.
To obtain TPB coatings, different procedures can be used. Evaporation and sputtering are the most common ways to obtain thin film deposition. Evaporation may be achieved by using a resistive heater or an electron beam heater. \\
The TPB coating of 8 in PMTs was performed for the first time in 74 ETL FLA9357 adopted by the ICARUS 
detector for the LNGS experimental phase. The depositions were obtained by spraying a
solution of TPB in toluene ($C_{7}H_{8}$) on the surface of each PMT\cite{GL}.
To obtain higher quality wavelength shifting, an improvement of this
technique is mandatory in view of forthcoming applications to cryogenic
detectors such as ICARUS and SBND within the SBN program at Fermilab and ProtoDUNE at CERN. \\
For this reason, the spray technique was replaced by an evaporation process, which allows to act on the coating parameters, such as the thickness, the quality and the uniformity of the layer, in a easier way. \\
Evaporation is a non-equilibrium process in which there is a net transport of liquid molecules across a liquid-vapour interface into the vapour phase.
The evaporation rate is described by the Hertz-Knudsen equation \cite{K-H}

\begin{equation}
\frac{1}{A}\frac{dN}{dt}=\frac{\alpha (P^{\ast}-P)}{\sqrt{2\pi mk_{B}T}}
\end{equation}
where $A$ is the area of the evaporating surface, $N$ is the number of molecules in the gas phase, $\alpha$ is the sticking coefficient of the gas molecules onto the surface ($0\leq\alpha\leq 1$), $P^{\ast}$ is the partial pressure of the gas in equilibrium with its condensed phase at a given temperature, $P$ is the ambient hydrostatic pressure acting upon the evaporant in the condensed phase, $m$ is the molecular mass, $k_{B}$ is the Boltzmann constant and $T$ is the temperature. 
Adopting a \textit{Knudsen cell} (K-cell), developed by Knudsen in 1909, $\alpha=1$ and the only two variables relevant for the evaporation process become the pressure $P$ and the temperature $T$. 
A K-cell consists of a crucible surrounded by an heating filament and a shutter on which an orifice is present.
A schematic design of the adopted K-cell is shown in figure \ref{K-cell}. \\
The basic idea to coat the PMT with TPB is to place it in front of a K-cell filled with a defined quantity of evaporating material. In this way the vapour, coming out from the cell lid orifice, hits the colder surface of the PMT window condensing in a thin uniform film. In the present case a thermal evaporation system was set-up by restoring a pre-existent apparatus consisting 
in a vacuum chamber instrumented with a K-cell.
However, since this system was used for different photodetectors with different size and shape, a careful study on the evaporation geometry
was required.   

\section{System geometry and simulation}

\begin{figure}
	\center
	\includegraphics[scale=0.35]{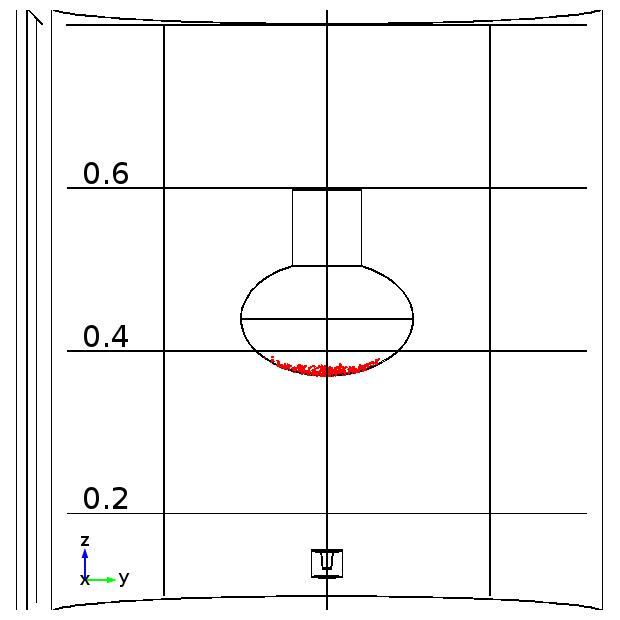} \hspace{1cm}
	\includegraphics[scale=0.35]{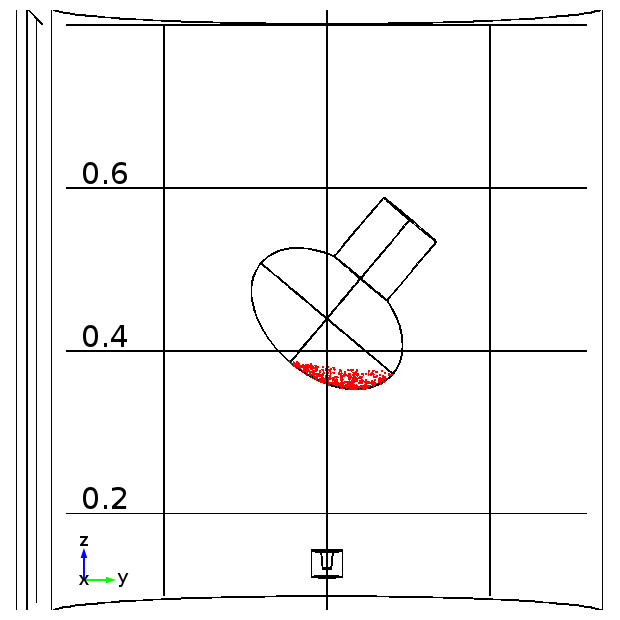}
	\caption{Graphical view of some adopted geometry for the COMSOL\textsuperscript{\textregistered}  particle tracing simulation. The resulting TPB particles distribution on the PMT sensitive windows at $0^{\circ}$ (\textit{left}) and $40^{\circ}$ (\textit{right}) orientation angles are shown.}\label{Comsol_spots}
\end{figure}

The assembly of the thermal evaporator 
required dedicated geometrical studies and system simulations carried out by means of 
\textit{COMSOL multiphysics}\textsuperscript{\textregistered}~\cite{COMSOL}.
Using this simulation tool, it was possible to design the most likely internal structure of the evaporator and to obtain some basic information correlated to the geometry of the evaporation process. 
The simulated structure is shown in figure \ref{Comsol_spots}. The PMT was placed above a K-cell at 30~cm distance with different orientation angles. This distance was determined by the PMT and vacuum chamber size, related to the available vacuum system (see section \ref{sec_system} for details).

\hspace{-0.74cm}The evaporation process was simulated by using the \textit{Particle Tracing} module supplied with COMSOL\textsuperscript{\textregistered}. In particular, the bottom of the crucible was considered a \textit{source} surface, emitting particles having a mass equal to that of TPB molecules, with isotropic direction (inside a 2$\pi$ semi-sphere), and with velocity given by the thermal distribution corresponding to the crucible temperature.
All the inner walls of the crucible were considered reflecting the particles
(having the same temperature of the bottom), while all the external surfaces (i.e. the PMT window
and the inner walls of the vacuum chamber), being at room temperature, were defined as \textit{freezing}
surfaces.
So the particles emitted from the bottom of the crucible bounced many times on its inner walls,
until they escaped from the small hole of the orifice, propagating without further interaction until
they reached the PMT cold surface. The angular distribution of the emitted particles was determined
by the geometrical characteristics of the crucible.
The particles reaching the PMT window were counted and their
positions were saved.
In this way, it was possible to calculate the amount of TPB deposited on the PMT window
and to determine its distribution on the surface.

\begin{figure}
	\center
	\includegraphics[scale=0.55]{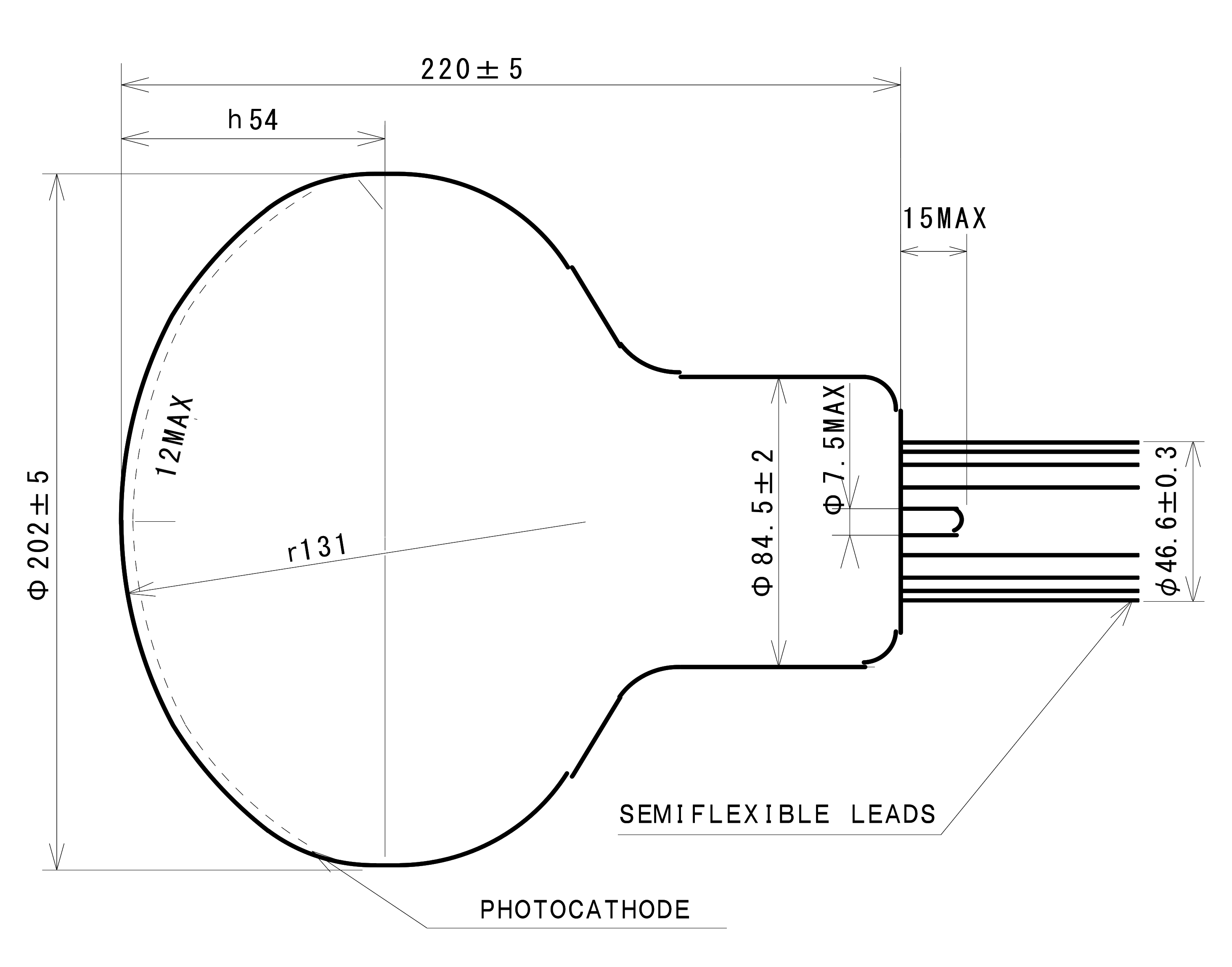}
	\caption{Technical drawing of the Hamamatsu R5912-MOD PMT.}\label{PMT}
\end{figure}

\begin{table}
	\center
	\begin{tabular}{c c }
		
		\hline
		
		PMT orientation ~~~~~ & Fraction of deposited \\
		angle & TPB (\textit{R}\/)\\
		
		\hline
		
		$0^{\circ}$ & $(42.6 \pm 0.2)~ 10^{-3}$ \\
		$10^{\circ}$ & $(44.4 \pm 0.2)~ 10^{-3}$\\
		$20^{\circ}$ & $(45.5 \pm 0.2)~ 10^{-3}$\\
		$30^{\circ}$ & $(51.1 \pm 0.2)~ 10^{-3}$\\
		$40^{\circ}$ & $(50.7 \pm 0.2)~ 10^{-3}$ \\
		
		\hline 
		
	\end{tabular}
	\caption{Fraction \textit{R} of deposited TPB on the PMT surface 
		with respect to the total evaporated quantity
		for different orientation angles. } \label{tab_angle}
\end{table}
\noindent As first, the simulation was used to evaluate the ratio between the number of TPB particles sticking on the PMT surface and the total number of particles coming out from the K-cell.
Assuming a 8~in PMT, whose geometry is shown in figure \ref{PMT}, the computation was performed for different angles of inclination of the PMT with respect to the K-cell axis.
The fraction \textit{R} of deposited TPB on the PMT surface with respect to the total evaporated quantity, 
is reported in table \ref{tab_angle}. Results show a slight increase of \textit{R} with the PMT inclination.
The PMT window area is approximately given by:

\begin{equation}
S=2\pi \cdot r \cdot h=41.9~cm^{2}.
\end{equation}
where $r=131$~mm is the radius of curvature of the sensitive area and $h=51$~mm its height, as shown in figure~\ref{PMT}. \\
Considering for examples the $40^{\circ}$ degrees orientation angle, the quantity of TPB needed to achieve a coating density of 0.2 mg/cm$^{2}$, assumed as target density value \cite{GL}, is:

\begin{equation}
\frac{0.2~mg/cm^{2} \times S}{R_{40^{\circ}}} = 1.65~g
\end{equation}
in agreement with past experimental results \cite{Burton}.

\begin{figure}
	\center
	\includegraphics[scale=0.29]{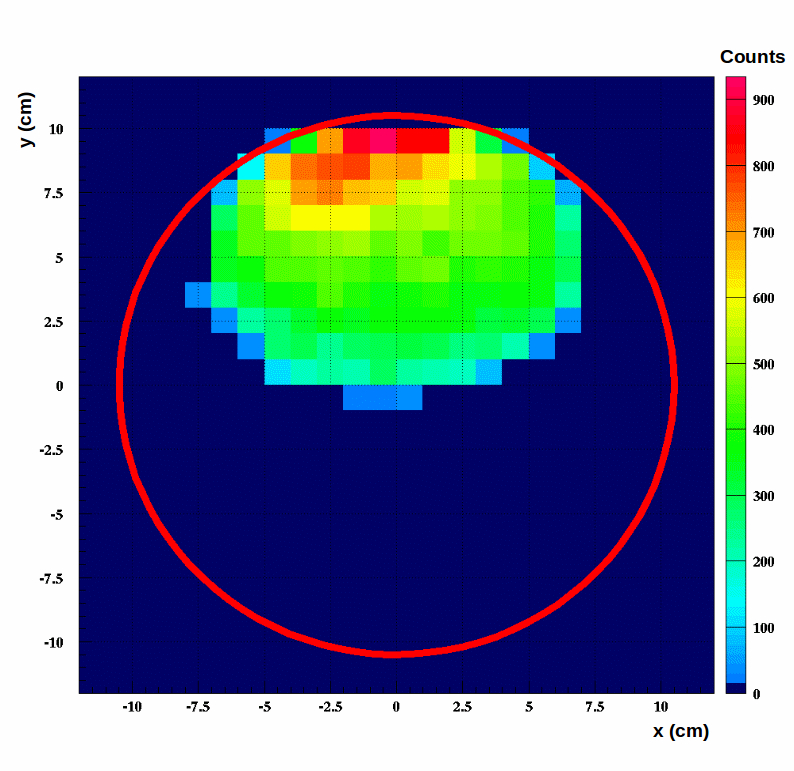} \hspace{1cm}
	\includegraphics[scale=0.29]{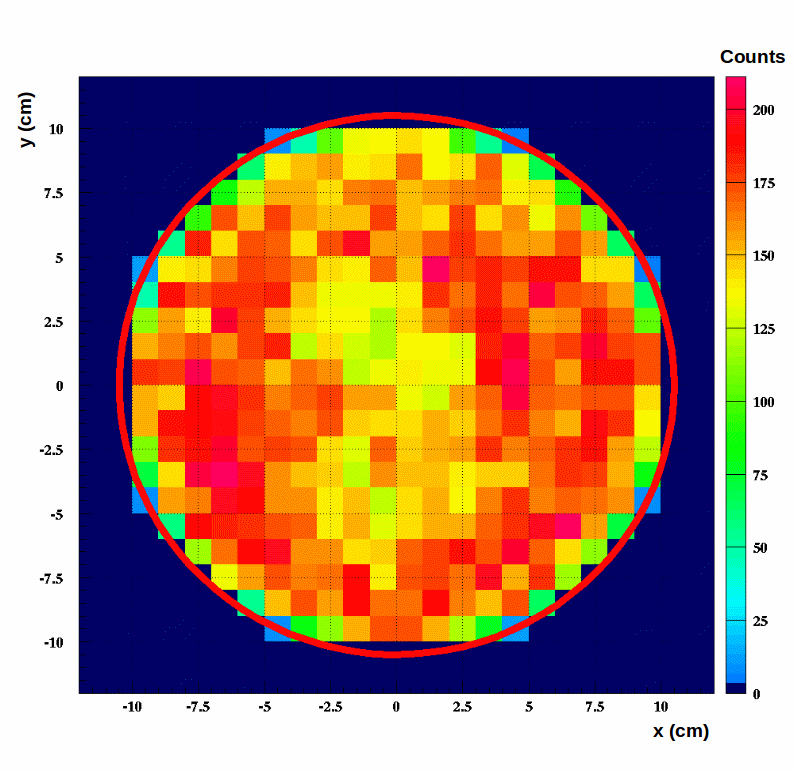}
	\caption{Graphical plots of the TPB particle distribution on the PMT sensitive
		window as resulting from the simulation.
		Distributions are shown for the PMT at $40^{\circ}$ orientation angle, 
		in fixed position (\textit{left}) and rotating on its own axis (\textit{right}).} \label{PMT_pizza}
\end{figure}

\begin{figure}
	\center
	\includegraphics[scale=0.7]{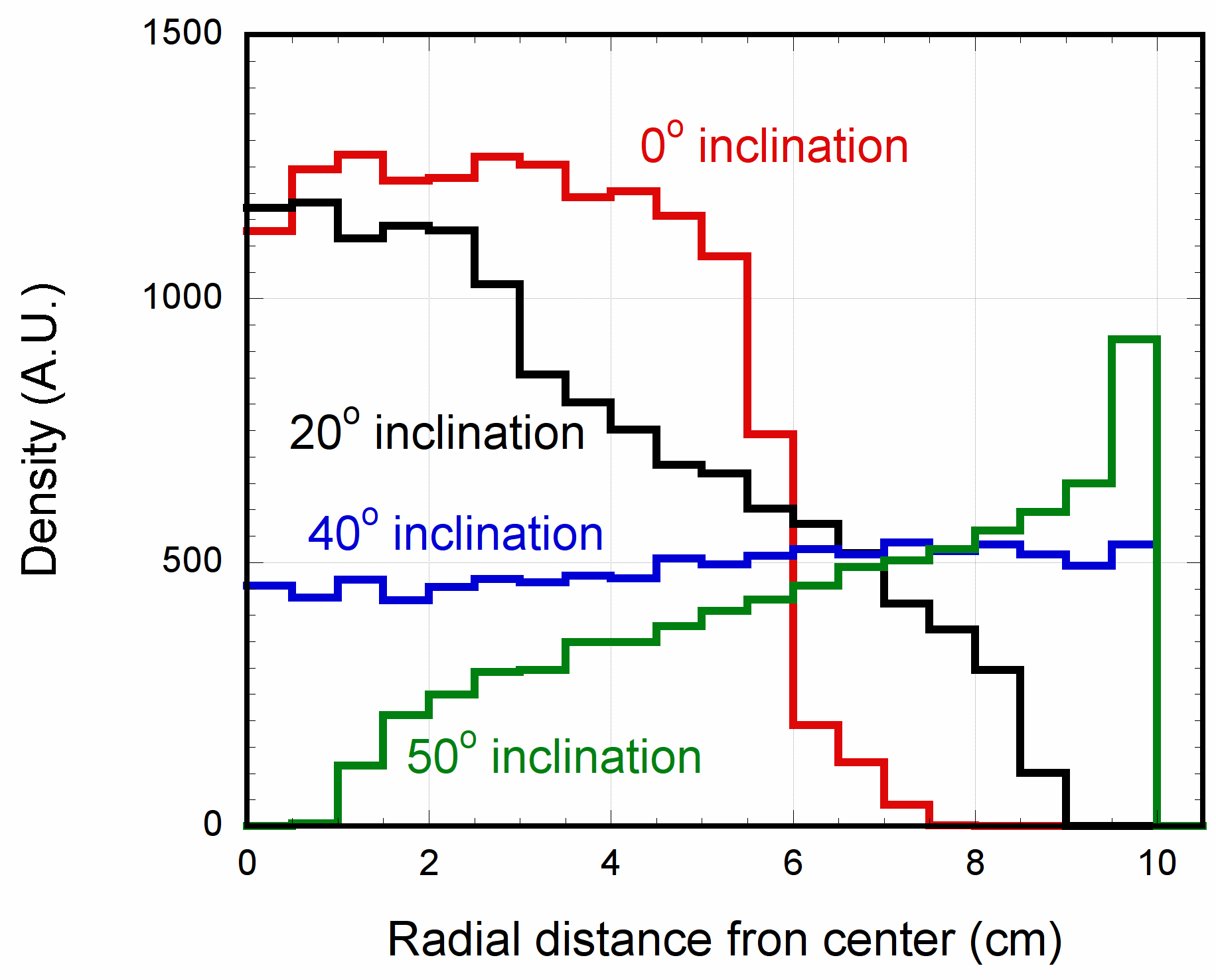}
	\caption{Plot of the TPB particle distribution as a function of the radial
		distance from the PMT window centre as resulting from the simulation. Curves are shown for the PMT oriented at $0^{\circ}$, $20^{\circ}$, $40^{\circ}$ and $50^{\circ}$ angle with respect to its own axis.} \label{TPB_angdis} 
\end{figure}

\noindent A further geometrical consideration that was taken into account
is the uniformity of the deposited layer, which mainly depends on the orientation of the PMT. 
Taking into account the $0^{\circ}$ orientation angle,  namely with the PMT axis aligned to the K-cell axis,
the geometrical analysis of the evaporation process showed that
the TPB deposition covers with a good uniformity only the central area of the PMT window, leaving 
totally uncoated the borders, as displayed in figure \ref{Comsol_spots} (left). 
Even for other orientation angles, 
the TPB covers with good uniformity only the portion of the sensitive window facing the K-cell, leaving the rest of the window almost totally uncoated, 
A more detailed analysis, shown in figure \ref{PMT_pizza} (\textit{left}), 
put in evidence that the resulting TPB density distribution for the PMT oriented at $40^{\circ}$ angle is negligible on most of the PMT window surface. \\
A considered solution was to apply a continuous rotation to the PMT around its own axis during the evaporation process, in order to distribute the TPB coating all over the sensitive surface. The effect of the PMT rotation during the evaporation process was simulated 
by rotating, in the PMT window reference plane, the coordinates of the TPB particle hitting the PMT.
As example, 
the resulting distribution of the TPB particles on the PMT window oriented at $40^{\circ}$ is presented in figure \ref{PMT_pizza} (\textit{right}), showing a good uniform distribution.
To find the optimal configuration, 
the TPB particle density was computed as a function of the radial distance from the PMT window centre supposing the device oriented 
with different angles ranging from $0^{\circ}$ to $50^{\circ}$. Some examples of resulting distribution are shown in figure \ref{TPB_angdis}.
As expected, in the $0^{\circ}$ configuration the TPB is distributed only in the central area of the PMT sensitive window, dropping sharply to zero at the edges. 
By increasing the orientation angle towards $40^{\circ}$, the TPB density tends to improve its uniformity all over the window surface, while its absolute value decreases in the central area. With orientation angles greater than $40^{\circ}$ the deposit is mainly on the outer region, with a lower density in the central one.\\
The $40^{\circ}$ PMT configuration was then considered as the best solution from the point of view of uniformity of the TPB coating.
On the basis of this result, the evaporation system was set-up by fixing the PMT inclined at an angle of $40^{\circ}$ with respect to the
evaporation axis.

\section{Evaporation system and procedure}
\label{sec_system} 

To perform TPB evaporations on a large number of 
PMTs, ensuring at the same time a high repeatability and reliability of the operation,
a dedicated thermal evaporator was set up and a specific evaporation
procedure was defined.

\subsection{The thermal evaporator}
\label{thick.moni}

\begin{figure}
	\center
	\includegraphics[scale=0.35]{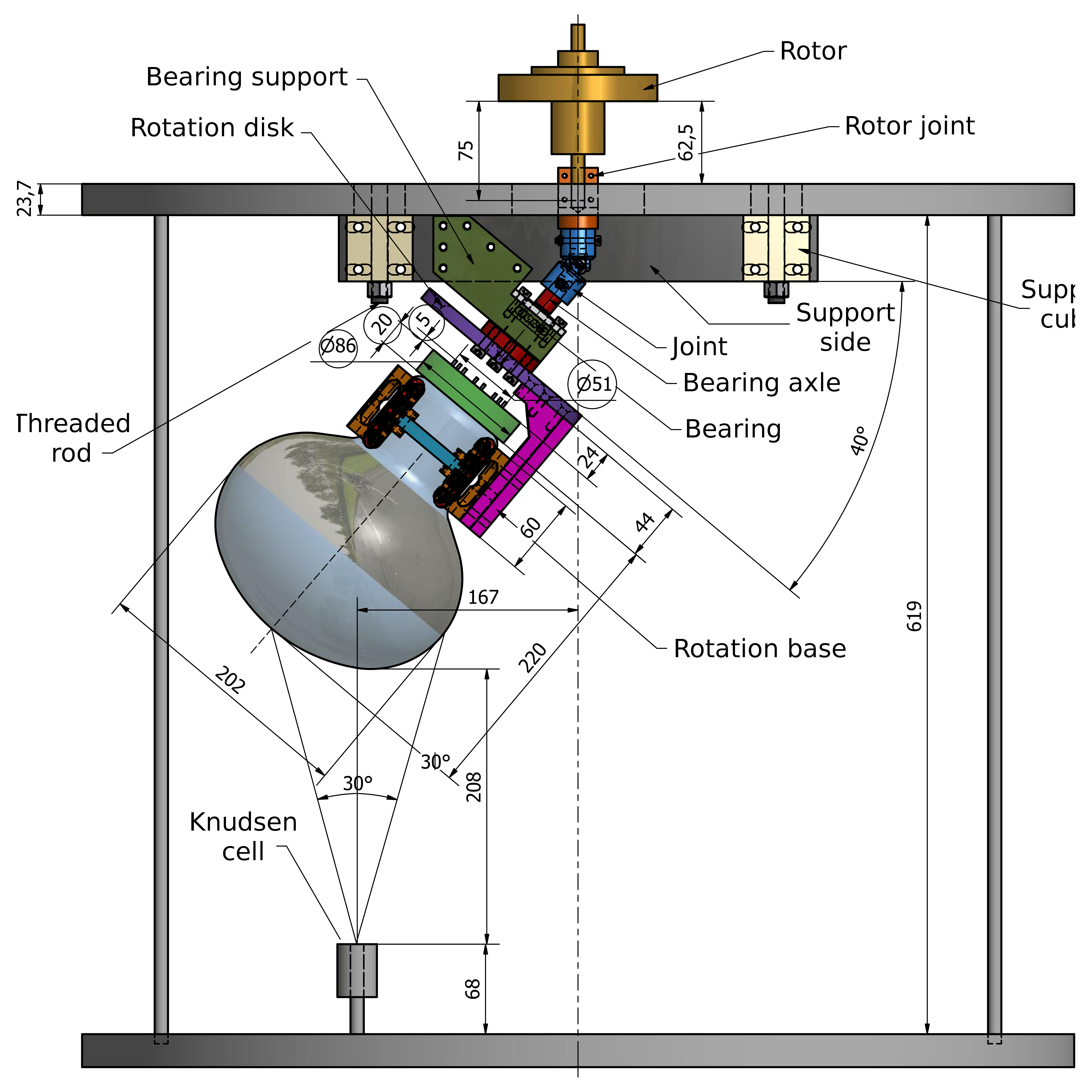} 
	\caption{Schematic design of the evaporation system.}
	\label{rot}
\end{figure}

The thermal evaporator consists of a vacuum chamber, 68~cm high and 64~cm diameter,
composed of two overlapping stainless steel cylinders. The chamber is closed at both sides 
by means of two large flanged plates.
The top plate constitutes the chamber cover and can be lifted up 
to allow access to the interior of the structure.
A schematic design of the evaporator is shown in figure \ref{rot}.

\noindent The vacuum chamber houses on the bottom plate 
a copper K-cell 
(see figure \ref{K-cell}).
The cell temperature is set by a \textit{Temperature Controller Ero Electronics LFS $1/16$ DIN} which allows to increase the cell temperature by an external heater. The cell temperature is measured
with a thermocouple located inside the K-cell. An external manual knob allows to rotate an internal shutter, placed above the K-cell,
used to intercept, 
when necessary, the evaporating particle flux. \\
\noindent The vacuum chamber is connected to two parallel pumping systems, used 
to obtain a preliminary vacuum grade and then to reach
the required high vacuum regime, respectively.
The first pumping system is composed of a single dry scroll pump (Varian TriScroll 600), which allows to achieve a pressure the order of $\sim 10^{-2}$~mbar. The secondary system consists of a rotary pump \textit{Edwards RV12} connected to a turbo-molecular pump \textit{Alcatel ATH 300C}, which is in turn connected to the chamber through a pneumatic valve that is opened when the pressure drops
below $10^{-1}$~mbar.
The internal pressure is measured by a \textit{Compact Full Range Gauge PKR251} and values are displayed on a \textit{Pfeiffer vacuum gauge controller}. \\
The PMT is fastened to a specific rotating holding support looking downwards and
inclined of $40^{\circ}$ angle
with respect to the vertical direction. 
The rotating structure, fixed  below the chamber cover,
is connected to an external 
motor (\textit{TRANSTECNO K80C63-7}) by a ferrofluid-based feedthrough allowing to keep the vacuum level during the PMT rotation. 
The whole rotating support allows rotation speeds up to
10~turns/min\\ 
During the evaporation, a \textit{Sycon STM-100/MF Thickness/Rate monitor} is used to have a feedback about the
evaporation process.
The monitor displays the growth of coating thickness both per time unit ($\angstrom$/s) and integrated on the whole process.

\subsection{The evaporation procedure}

\begin{figure}
	\center
	\includegraphics[scale=0.6]{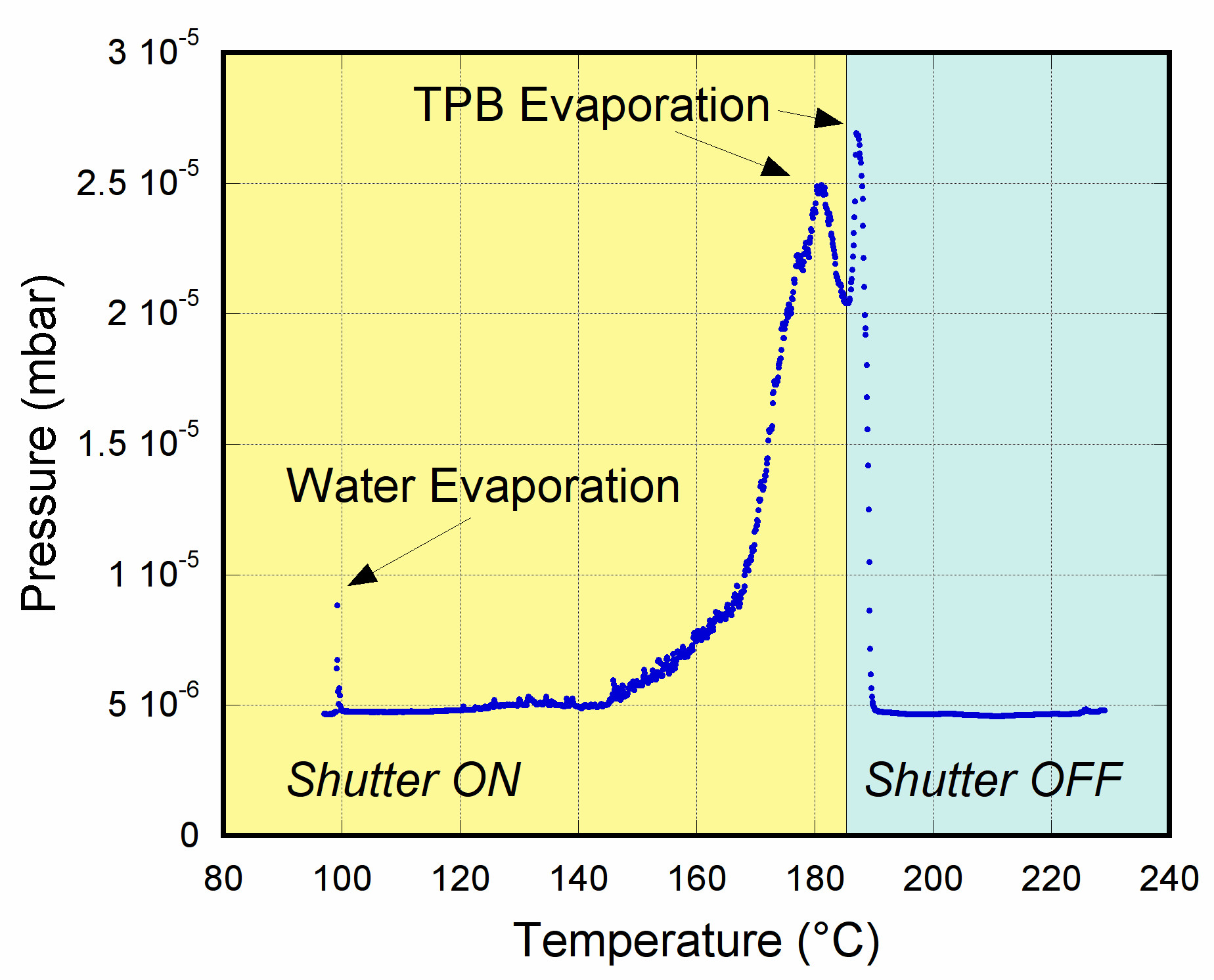} 
	\caption{
		Pressure value during the whole evaporation process. 
		As the temperature increases, it is possible to observe in order: a peak produced by water evaporation;
		a peak corresponding to the TPB boiling point; a peak in coincidence with the shutter opening.} 
	\label{pressure}
\end{figure}

The vacuum chamber is opened by lifting up the top plate 
and a clean PMT is fastened to its holding support. The 
K-cells crucible is filled with the defined amount of TPB.
After closing the chamber
the vacuum pumping procedure is started. 
The evaporation process can start when the pressure inside the chambers falls below $10^{-5}$~mbar. 
The K-cell temperature heater is set to 
$220^{\circ}$C (see Section~\ref{Rate}) and the PMT rotation in switched on. At $100^{\circ}$C the internal pressure undergoes brief variations due to
the evaporation of water and other impurities condensed on the cell, as shown in figure \ref{pressure}. At $\sim 185^{\circ}$C TPB reaches its boiling point and a second peak of pressure is observed. Until this moment the shutter is kept closed in order to avoid that water and other impurities reach the surface of the PMT. When pressure drops back, the shutter can be opened and the real evaporation process starts. At this moment, pressure undergoes a last peak before dropping down again to few $10^{-6}$~mbar. The growth of the TPB deposition is continuously checked by the thickness monitor, as described in Section~\ref{Rate}.
When all the material contained in the cell has evaporated, the thickness growth drops to zero and the process can be considered concluded. After that, the system is kept at vacuum condition for at least $\sim 20$~min in order to allow the stabilization of the TPB condensate layer on the PMT surface and the lowering of the temperature below $100^{\circ}$C. The vacuum is then broken by means of nitrogen venting.

\section{Measurement of the deposited TPB density and uniformity}\label{3Dprinted}

\begin{figure}
	\center
	\includegraphics[scale=0.06]{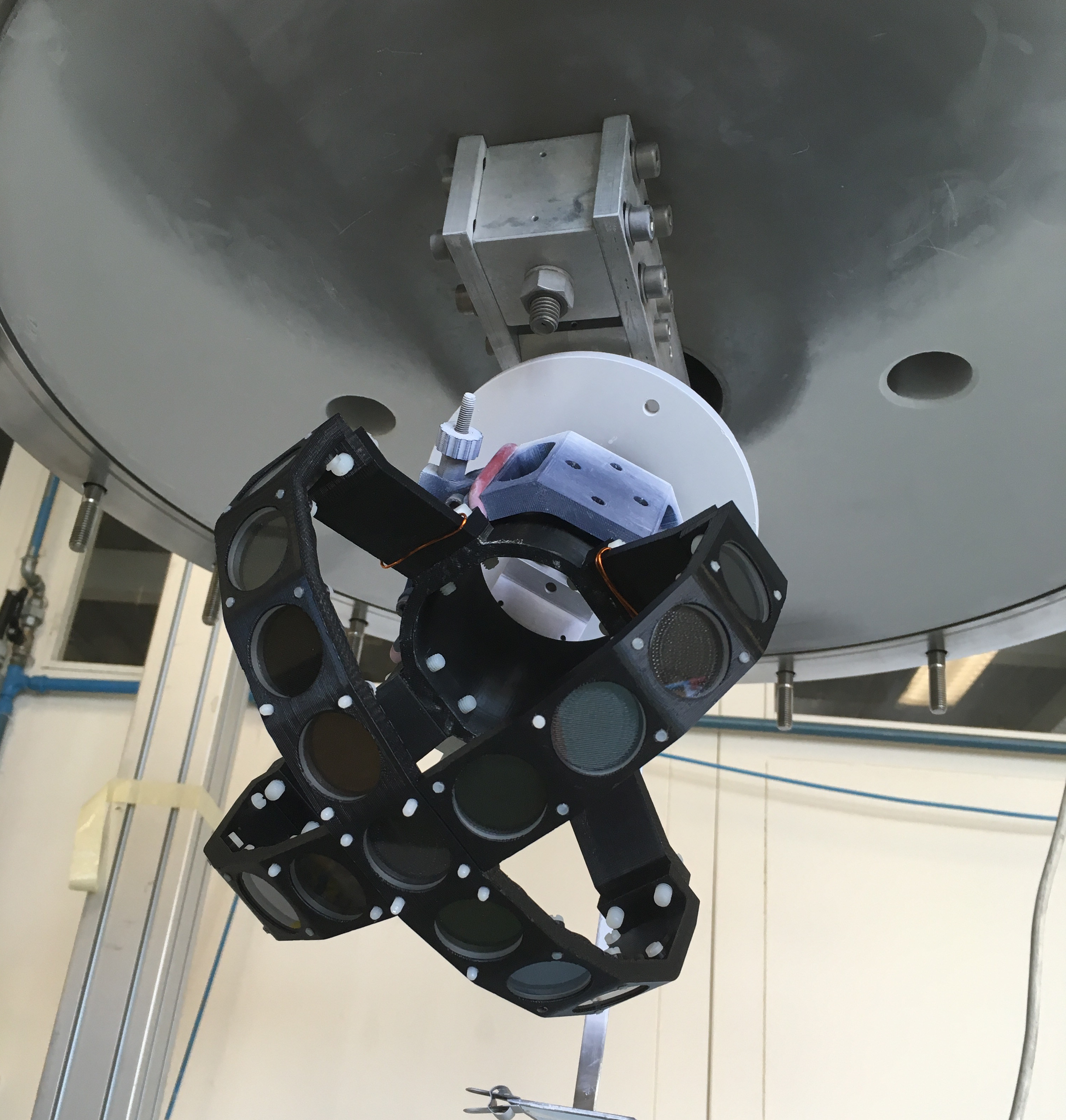} \hspace{1cm}
	\includegraphics[scale=0.6]{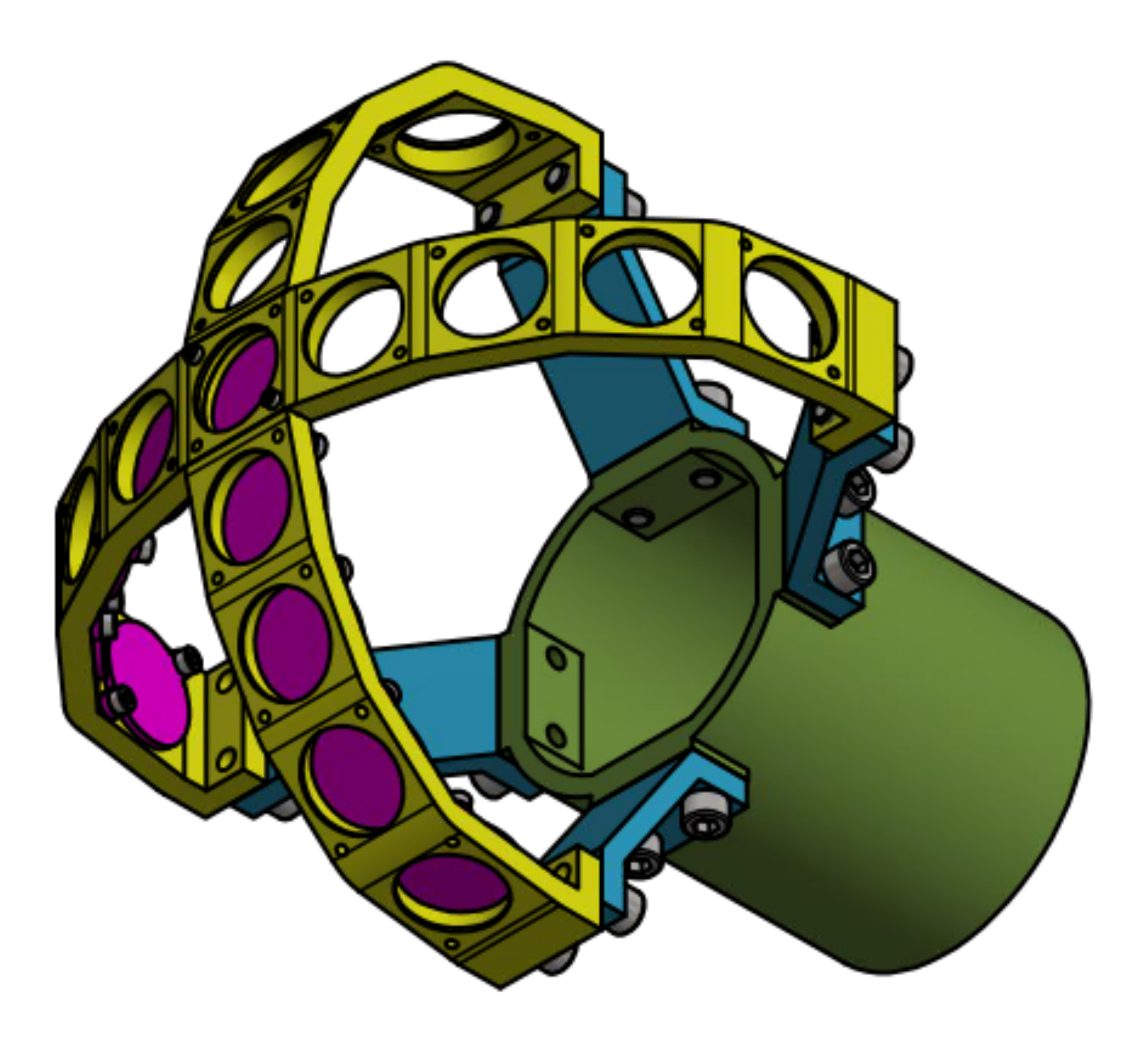}
	\caption{Picture (left) and CAD drawing (right) of the 3D-printed support which allows to perform density measurements by using round glass slides.
		Slides are visible in their locations.}
	\label{new_support} 
\end{figure}

The effectiveness of the rotating support system was verified, as first, by checking the deposited TPB density and uniformity
on the PMT sensitive window. 
A direct measurement on PMTs with curved and sandblasted windows is not trivial. 
The adopted solution was the realization of a PMT mock-up allowing to measure the local density in different positions on the PMT surface by means of round glass slides in different positions.
The mock-up consists of 
a plastic support with the same radius of curvature of the PMT, designed and realized by means of a 3D printer (figure \ref{new_support}). It supports up to 
$17$ round glass slides of 1~in diameter inside 
adequately positioned flared holes.
The geometry consists of a central slide plus 16 slides symmetrically deployed
on 4 arms along two orthogonal axes, according the following position table:

\begin{center}
	\begin{tabular}{c c c c}
		\hline
		Position & Location & Radial distance  & Number \\
		&          & from centre      &  of slides            \\
		\hline
		$P_0$ & Central & 0 cm & 1 \\
		$P_1$ & First & 3 cm & 4 \\
		$P_2$ & Second & 6 cm & 4 \\
		$P_3$ & Third & 9 cm & 4 \\
		$P_4$ & Fourth & 12 cm & 4 \\
		\hline
	\end{tabular}
\end{center}

\noindent The weight of the each slide was measured before and after the evaporation process by means of a balance with an accuracy of $10^{-4}$~g. 
Evaporation was carried out with $0.8$~g of TPB.



\begin{figure}
	\center
	\includegraphics[scale=0.65]{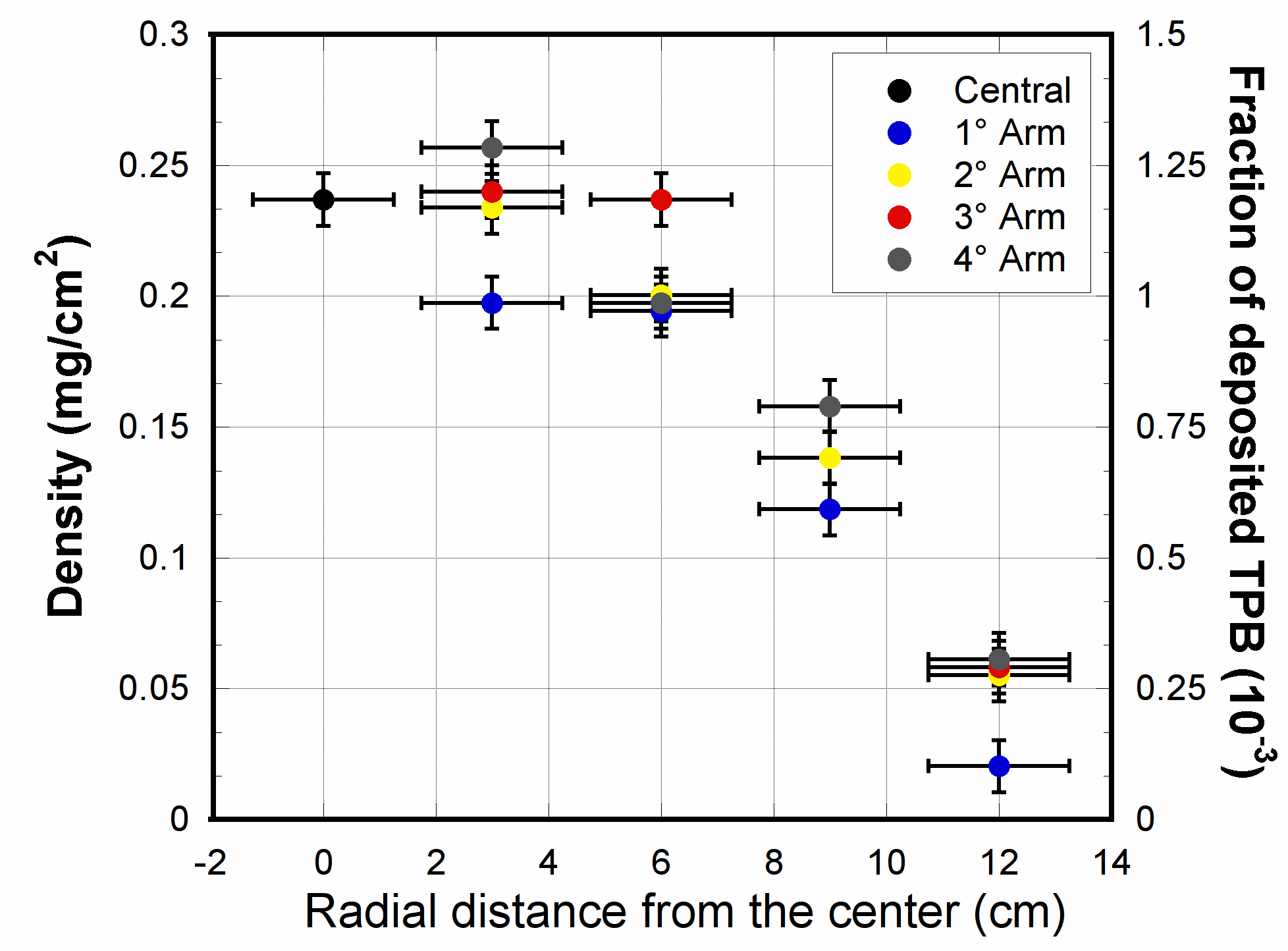} 
	\caption{Resulting densities for all the $17$ slides as a function of the radial distance from the centre
The corresponding fraction of the whole used TPB deposited per slide is also shown.}\label{den_slides}
\end{figure}

\noindent The resulting density values, presented in figure \ref{den_slides}, show a good uniformity ($\pm 15\%$) within the central slide ($P_0$) and the slides located in the first two positions along the 4 arms
($P_1$ and $P_2$), spanning a radial distance from the centre up to 7~cm. A decrease of about 50\%
is evident in the slides located in the position $P_3$, at a mean radial distance from the centre of 9~cm.
On position $P_4$ the TPB density drops by a factor 10, mainly due to border effects since the slide area extends outside
the evaporation cone.

\section{Measurement of the Quantum Efficiency}\label{QE_slides}

In order to verify the reliability of evaporation system and coating process in terms of light conversion efficiency, an optical test system was set up to allow the measurement of the Quantum Efficiency (QE) of a generic photodetector (PMT, SiPM, photodiodes...),
i.e. the probability that an impinging photon will produce a photoelectron in the device.
%
The overall QE of a photodetector at a given wavelength $\lambda$
can be experimentally evaluated comparing the currents given by the photodetector under test and by a reference calibrated photodiode, according to the equation:

\begin{equation}
QE_{K}(\lambda)=QE_{D}(\lambda)\times \frac{I_{K}(\lambda)}{I_{D}(\lambda)}
\end{equation}
where $QE_{D}(\lambda)$ is the tabulated value of the reference photodiode QE as a function of the wavelength $\lambda$ of the incident light and $I_{K}(\lambda)$ and $I_{D}(\lambda)$ are the photodetector and photodiode cathodic currents,  respectively. 
In the VUV region, this measurement has to be realized
in vacuum by focusing a monochromatic source on the calibrated photodiode at first, and then on the photodetector under test.
The currents resulting from both the photodiode and the photodetector under test are alternately measured by a pico ammeter

\subsection{Set-up of the optical test system}\label{monochromator}

\begin{figure}
	\center
	\includegraphics[scale=0.2]{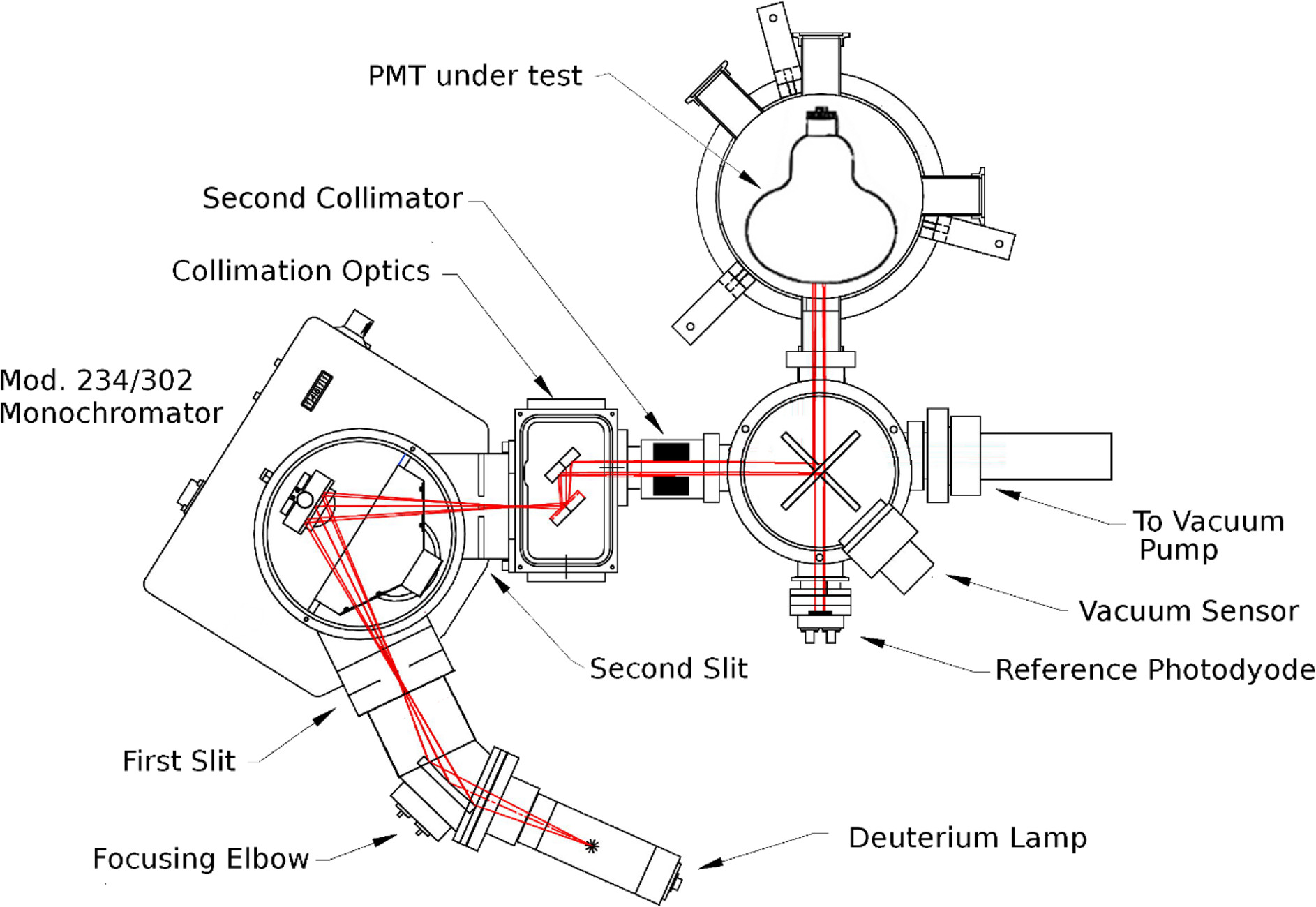} 
	\caption{Experimental set-up of the optical test system used for the Quantum Efficiency measurements.} \label{optical}
\end{figure}

\begin{figure}
	\center
	\includegraphics[scale=0.65]{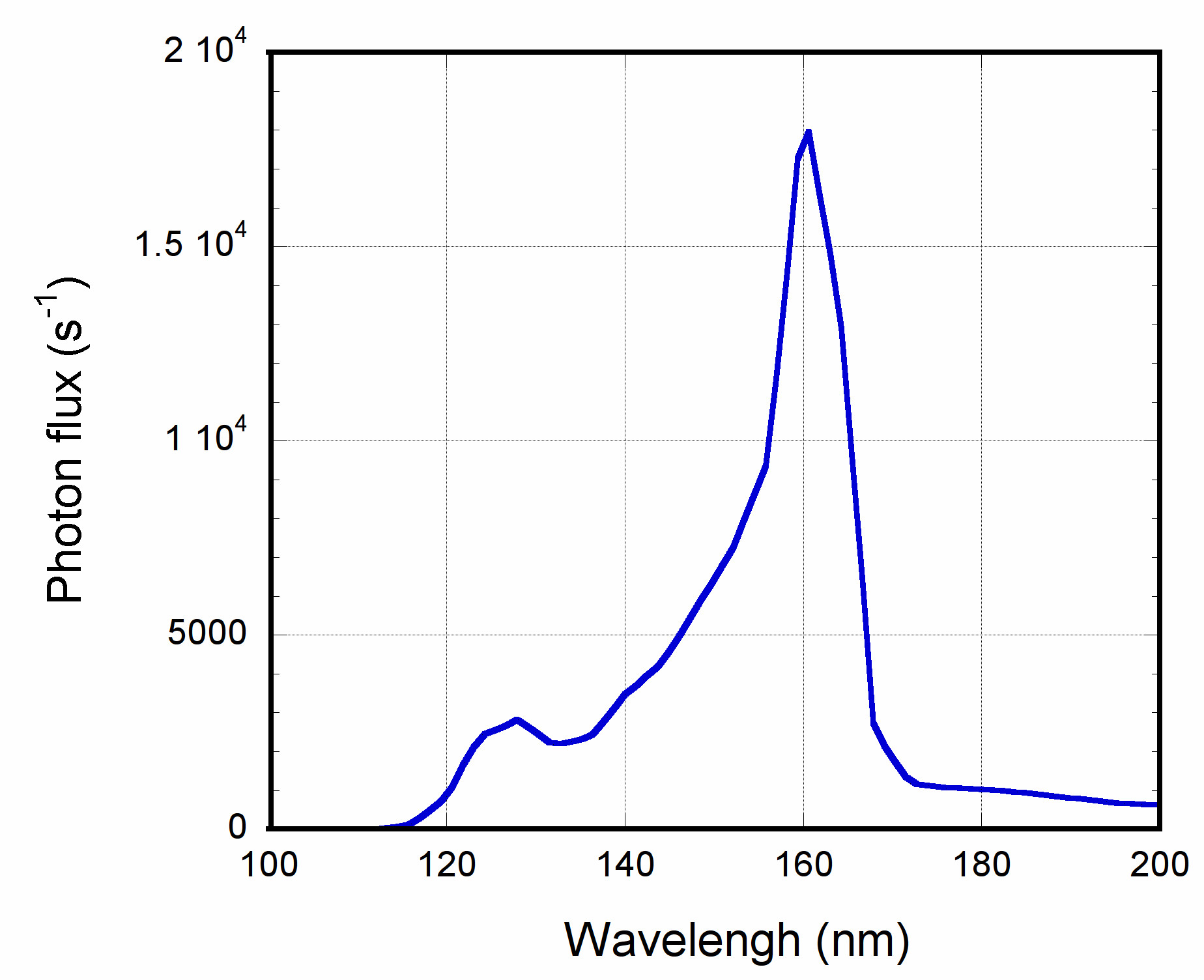} 
	\caption{Resulting spectrum of the light source at the monochromator output.
		Two peaks are present at 128~nm and 160~nm.} \label{d2spectrum}
\end{figure}

The optical test system used for the QE measurement is shown in figure \ref{optical}. 
It consists of a light source, a monochromator, a focusing system and a chamber housing the device under test.
A turbo-molecular pumping system allows the achievement of ultra high vacuum regime on the whole apparatus
(few $10^{-5}$~mbar) during the measurement. \\
The light source is a 30~W deuterium lamp (McPherson 632)
whose emission spectrum ranges from $115$~nm to $380$~nm. 
A focusing elbow and a UV-reflective mirror allow to send the light to a monochromator (McPherson 234/302). The light enters into the monochromator through a first slit, which allows to calibrate its intensity. A diffraction grating ($120~$ridges/mm) and a second slit allow to select the light wavelength and to regulate the range of its amplitude.
The monochromator scan is controlled by a computer which allows to select a specific wavelength or to scan
the whole lamp spectrum with a settable pitch. A double collimation system
allows to stabilizes the light beam size and to reduce its divergence. 
The light beam is sent towards a rotating mirror which allows to selectively direct it towards the calibrated photodiode or to the photodetector under test. 
Figure \ref{d2spectrum} shows the resulting photon wavelength spectrum at the monochromator output.
The spectrum shows two peaks around 128~nm and 160~nm that can be considered as the wavelengths to be used for the tests.

\begin{figure}
	\center
	\includegraphics[width=6cm]{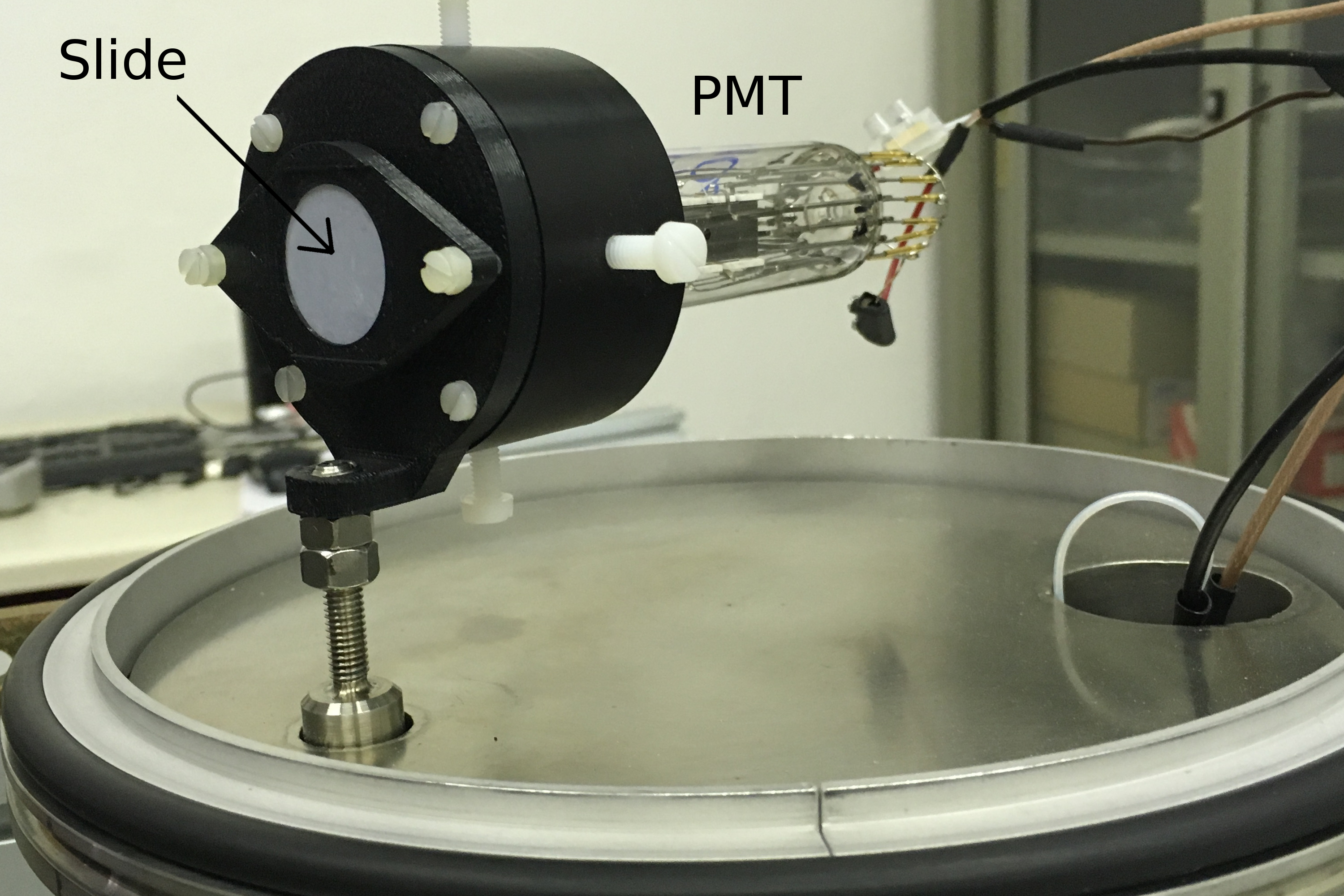} \hspace{0.5cm}
	\includegraphics[width=6cm]{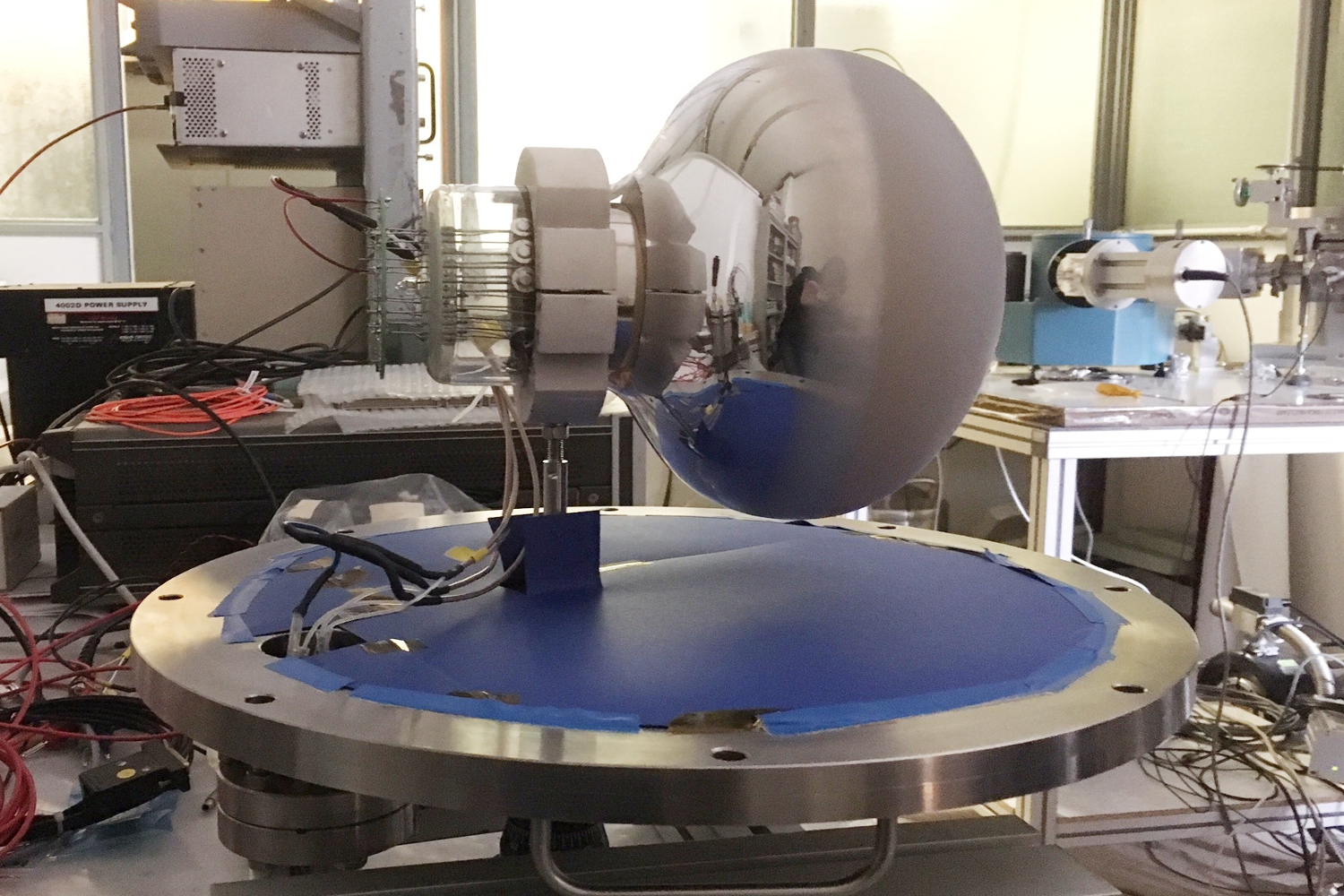} 
	\caption{Photodetector support for quantum efficiency measurements which allows the transverse rotation of the device.
		Slides were placed in front of a 1.5~in photomultiplier by means of the black plastic support shown in picture on the left.
		On the right a 8~in PMT is mounted on the support.}\label{glass_supp}
\end{figure}

\noindent The photodetector under test is housed in a dedicated chamber by means of a mechanical support which allows the transverse rotation
of the device. For slide measurement, glasses were placed in front of a 1.5~in photomultiplier \textit{ETL D830} by means of a plastic support as shown in figure \ref{glass_supp}. 
The chamber size (30~cm diameter) is sufficiently large to allow the housing of a 8~in diameter PMT and permit the evaluation
of the effective QE of the device.
PMT was used as a diode, with the multiplication stage shorted at $+200~V$ and the photo cathode grounded through the pico ammeter Measurements were performed by setting the monochromator entrance slit at 3~mm aperture, while the output slit was set at 1~mm aperture. The resulting wavelength resolution of the outgoing light was of the order of 4~nm. 


\subsection{Quantum Efficiency vs slide position}

\begin{figure}
	\center
	\includegraphics[scale=0.65]{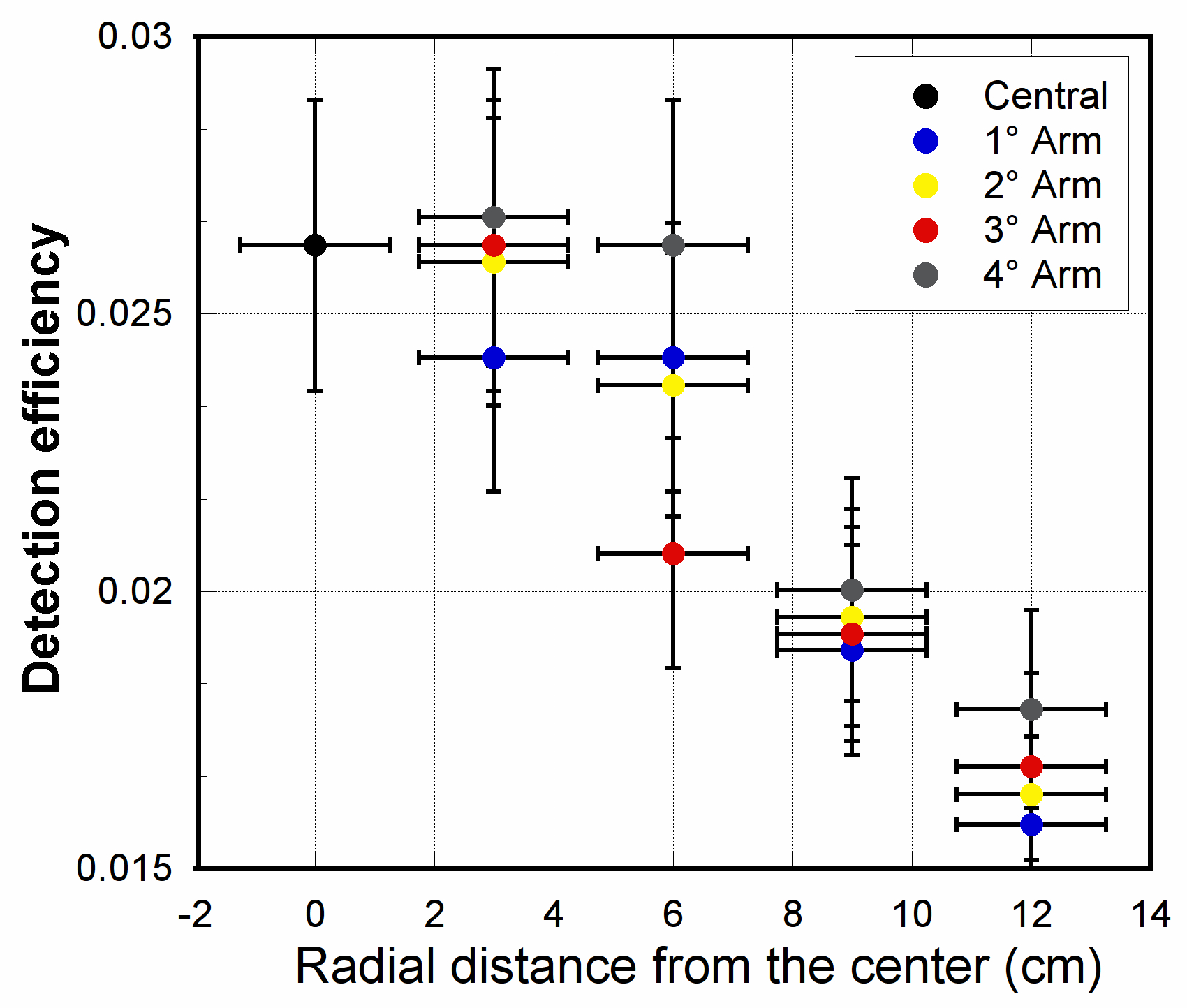}
	\caption{Conversion efficiency for the 17 slides as a function of their location on the 3D-printed support.}\label{QE_glass} 
\end{figure}

Round glass slides were at first used to evaluate the coating uniformity on the PMT surface from the point of view of the light conversion
efficiency. Slides were coated in different mock-up positions starting with an initial TPB quantity of $\sim 0.8$~g.
To measure the QE, the method already described above, was followed.
Results are reported in figure \ref{QE_glass}.
They show a good uniformity of the light conversion efficiency, in agreement with the previous results 
(see Section \ref{3Dprinted}). The conversion efficiency also shows a good uniformity between the centre ($P_0$) and the 8 slides placed in the positions $P_1$  and $P_2$. A $40\%$ reduction of the QE is observed in the farther positions $P_3$ and $P_4$. 

\subsection{Quantum Efficiency vs evaporation rate}\label{Rate}

\begin{figure}
	\center
	\includegraphics[scale=0.6]{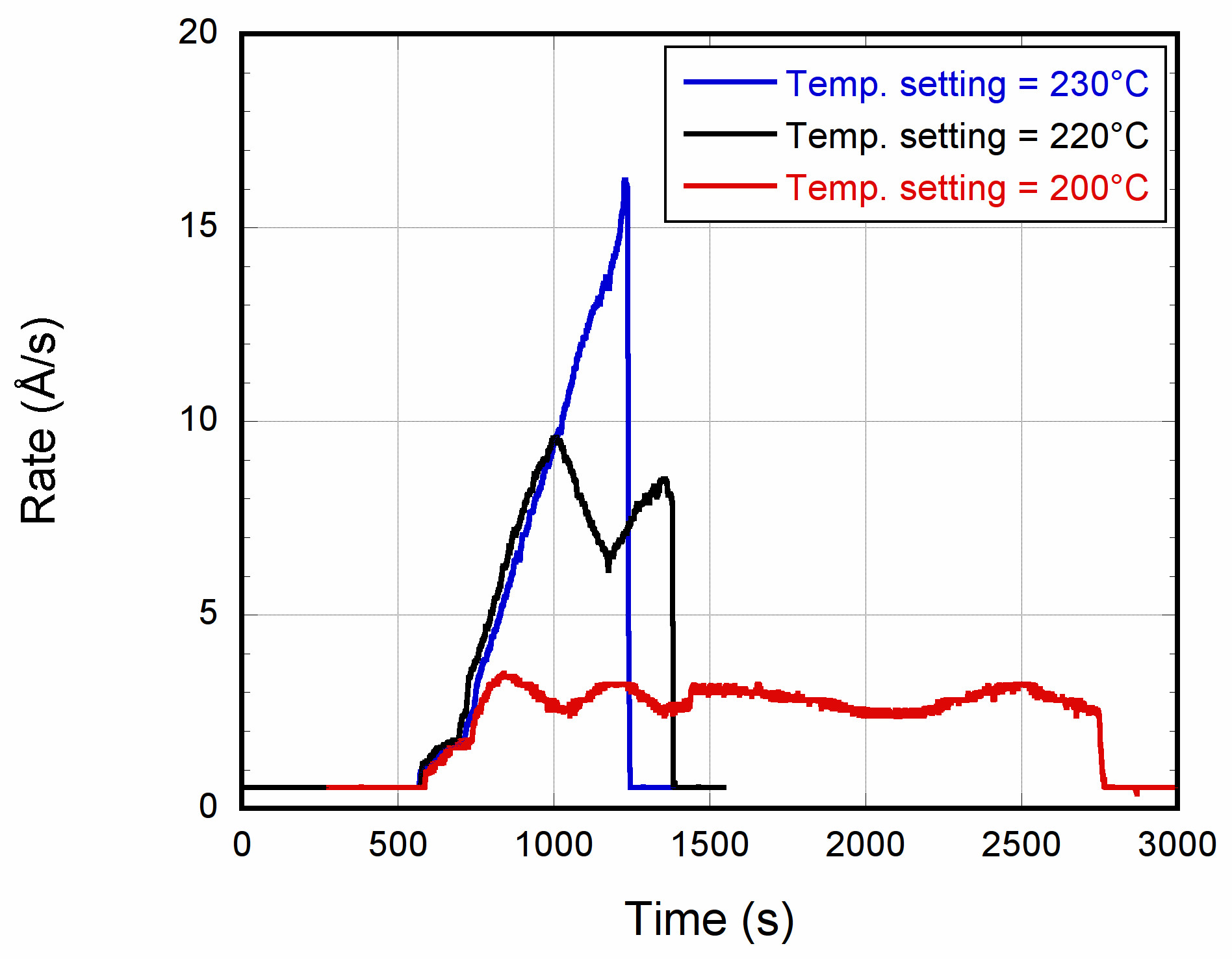}
	\caption{Evaporation rate for three different evaporation procedures.}\label{rate} 
\end{figure}

The thickness monitor installed in the evaporation system (see Section \ref{thick.moni}) allows to observe the deposition growth during the evaporation process. In particular, this device displays the total thickness of the layer deposited on the crystal (measured in k$\angstrom$ units) and the evaporation rate, namely the coating thickness per unit of time (measured in $\angstrom$/s units). This last parameter depends on the temperature set during the process. A series of tests was performed in order to identify a possible correlation between the evaporation rate and the quality of the deposition in terms of QE. \\
A set of slides was coated using processes with three different values of the evaporation rate.
The resulting rate curves are shown in figure \ref{rate}.
In the first process, the heater temperature was set to $200^{\circ}$C and left at this value for all the duration of the process, which took about $40$~min; the rate showed some fluctuations,
due to the heater on-off cycle, around the mean value of $2\div3~\angstrom$/s. 
In the second case, the heater temperature was set to $220^{\circ}$C and the evaporation process took $\sim 15$~min; the displayed average rate was around $8\div9~\angstrom$/s and lasted
for two on-off cycles of the heater.
Lastly, a third evaporation was carried out by setting the heater temperature to $230^{\circ}$C; the evaporation process took few minutes and the displayed average rate was around $15\div16~\angstrom$/s. All the three evaporation processes were carried out starting with the same initial TPB quantity of $0.8$~g and the final total thicknesses on the crystal resulted in $\sim 4$~k$\angstrom$.\\
Slides were tested by the described optical system.
The resulting mean conversion efficiencies as a function of the TPB coating density are shown in Tab.
\ref{tab_rate}. 

\begin{table}
	\center
	\begin{tabular}{c c}
		\hline
		Evaporation rate ($\angstrom$/s) & Conversion efficiency \\
		$\angstrom$/s                    & Relative units        \\
		\hline
		$2\div3$                & $0.025 \pm 0.001$ \\
		$8\div9$                & $0.027 \pm 0.001$ \\
		$15\div16$              & $0.029 \pm 0.001$ \\
		\hline 
	\end{tabular}
	\caption{Mean value of conversion efficiency
		as a function of the coating density as resulting for three sets of slides coated at different 
		evaporation rate value.} \label{tab_rate}  
\end{table}
\noindent Data suggest a possible correlation between conversion efficiency and evaporation rate,
probably due to the presence of residual impurities in the chamber which mostly affect the slowest evaporation processes. 
However, as shown in figure \ref{rate}, the fastest evaporation process is extremely difficult to control, due to its short duration 
which took only few minutes. 
Taking into consideration that 
a fast evaporation on a rotating target could affect the TPB coating in terms of uniformity,
an evaporation process with a mean rate of $8\div9~\angstrom$/s, corresponding to  
a heater temperature set of $220^{\circ}$C, 
was considered as the most suitable for the evaporation procedure.

\subsection{Quantum Efficiency measurements on PMTs}

\begin{figure}
	\center
	\includegraphics[scale=0.65]{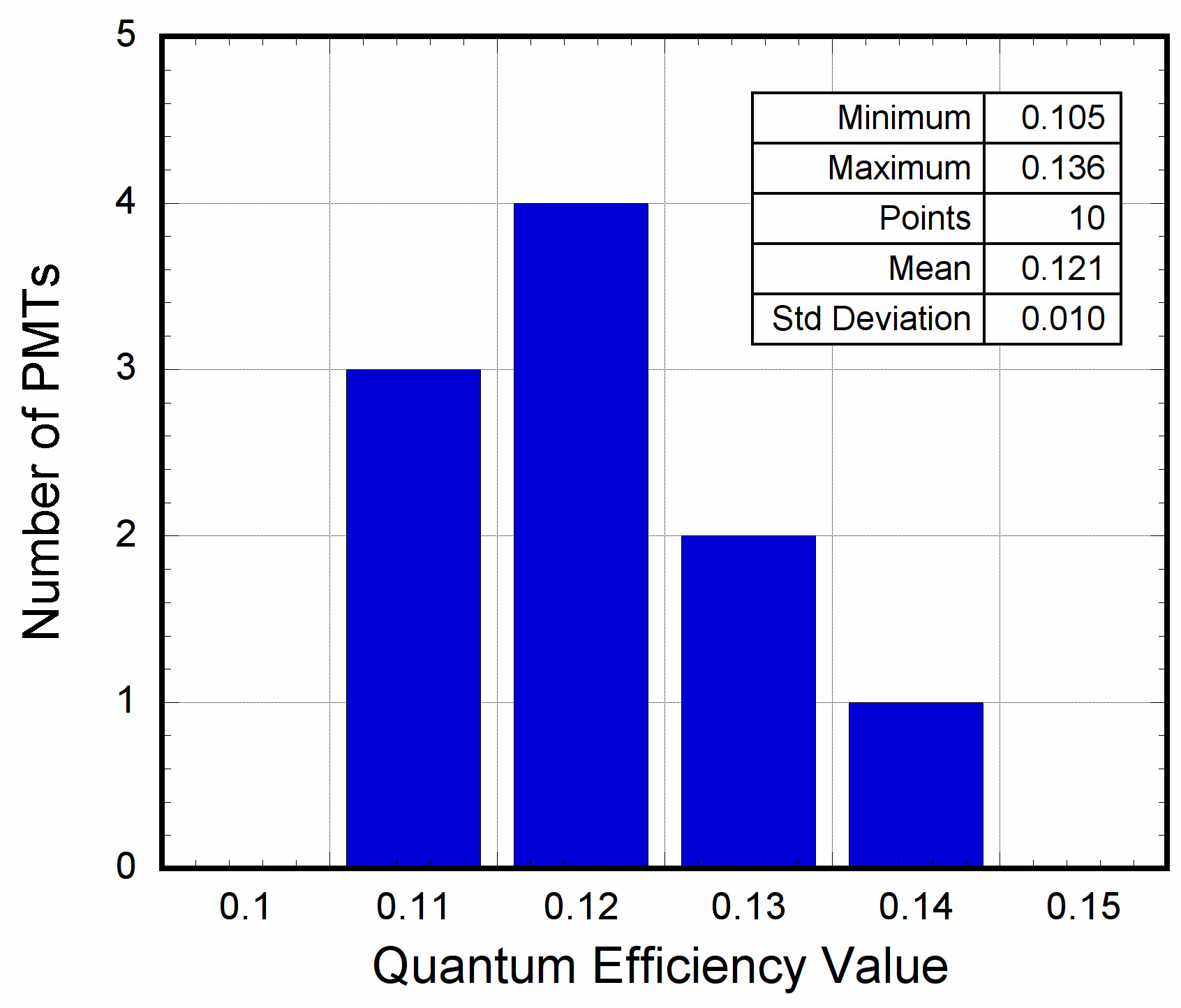}
	\caption{Distribution of the quantum efficiency over the tested PMTs.}
	\label{distr_qe} 
\end{figure}

Before proceeding with the coating of all the required PMTs, a final test was carried out
by performing the evaporation and direct measurement on a small sample of 8~in PMTs.
A set of 10 Hamamatsu R5912-MOD PMTs was coated following the standard evaporation procedure 
described above.
For each device, the absolute QE was evaluated by illuminating 
the window centre at $\lambda=128$~nm using the same optical test system described
in Section~\ref{monochromator}. PMT was used as a diode, with the multiplication stage 
in shorted at $+200~V$ and the photo cathode directly grounded through the pico ammeter \\
The resulting QE values are shown in figure \ref{distr_qe}, presenting values
ranging from 0.11 to 0.15, with 0.12 as average value.

\begin{figure}
	\center
	\includegraphics[scale=0.65]{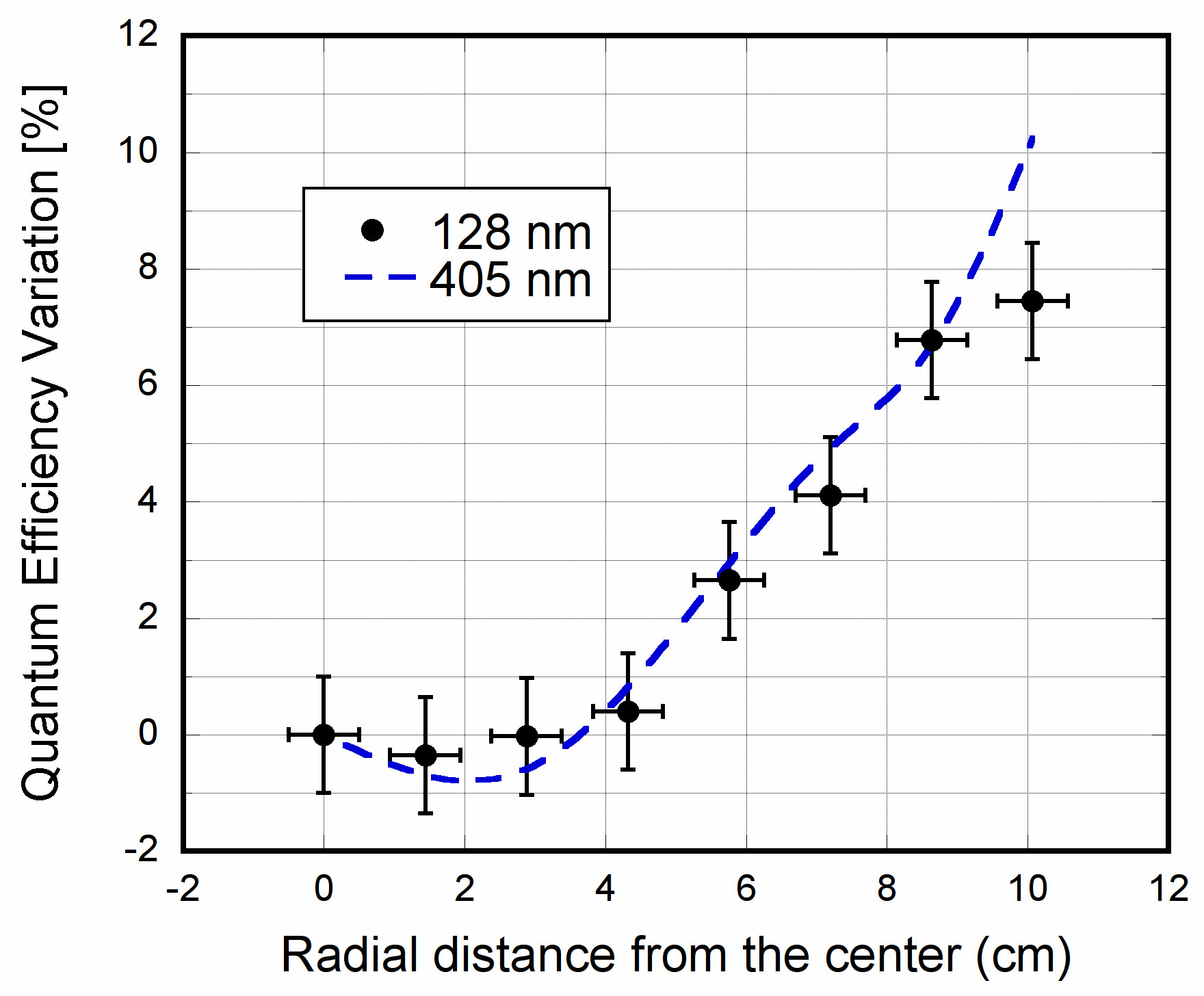}
	\caption{Variation of the QE as a function of
		the radial distance from the window centre. The dashed curve represents the sensitivity variation of the bialkali photo cathode to visible light.}
	\label{plot_qe} 
\end{figure}

\noindent Measurements were also repeated by illuminating the PMT window in different position,
by rotating the PMT inside the housing chamber by means of an external knob.
Figure \ref{plot_qe} shows the QE trend as a function of
the radial distance from the window centre. Starting from the centre, it
can be noted a slight decrease of the QE, followed by a continuous increase up to
the windows border. These variations, within $\pm 5\%$ around the QE mean value,
reflect the sensitivity variation of the bialkali photo cathode to visible light
without TPB coating, as shown by
the dashed curve in the same plot~\cite{R5912}\cite{R5912_B}.



\section{Conclusions}

A specific evaporation technique for the deposition of Tetraphenyl-butadiene on the surface
of 8~in convex window photomultiplier tubes was developed. A dedicated evaporation system, which make use of a \textit{Knudsen cell}
and a rotating sample support, was set up.
Simulation results and experimental tests demonstrate the effectiveness of this technique from the point of view of deposition uniformity and light conversion efficiency. \\
The outcomes presented in this work validate this technique for series deposition of a large
number of PMTs and it was successfully adopted for the 360 Hamamatsu R5912 of the upgraded 
ICARUS T600 light detection system.

\section*{Acknowledgement}

This study was carried out in the framework of the CERN Neutrino Platform NP01 activities.
The thermal evaporator and the optical test system described in this paper were funded
by the italian INFN (``Istituto Nazionale di Fisica Nucleare'') and MIUR 
(``Ministero dell'Istruzione, dell'Universit\`{a} e della Ricerca'') within the PRIN
(``Progetto di Rilevante Interesse Nazionale'') program.
The authors of this paper thank the CERN Technology Department staff for hosting the  thermal evaporator and for the fruitful technical collaboration and discussion.

\end{document}